\documentclass[reprint,twocolumn,showpacs,preprintnumbers,amsmath,amssymb,superscriptaddress]{revtex4}
\usepackage{graphicx}
\usepackage[hypertex]{hyperref}

\begin{document}

\bibliographystyle{prsty}
\title{Angle-resolved photoemission spectroscopy study of PrFeAsO$_{0.7}$:\\
Dependence of the electronic structure on the pnictogen height}

\author{I. Nishi}
\email{nishi@wyvern.phys.s.u-tokyo.ac.jp}
\affiliation{Department of Physics and Department of Complexity 
Science and Engineering, University of Tokyo, 
Hongo, Tokyo 113-0033, Japan}

\author{M. Ishikado}
\affiliation{Japan Atomic Energy Agency, Tokai, Ibaraki 319-1195, Japan}
\affiliation{Nanoelectronic Research Institute, National Institute of Advanced Industrial Science and Technology (AIST), Tsukuba, Ibaraki 305-8568, Japan}
\affiliation{JST, TRIP, Chiyoda, Tokyo 102-0075, Japan}

\author{S. Ideta}
\affiliation{Department of Physics and Department of Complexity 
Science and Engineering, University of Tokyo, 
Hongo, Tokyo 113-0033, Japan}

\author{W. Malaeb}
\affiliation{Department of Physics and Department of Complexity 
Science and Engineering, University of Tokyo, 
Hongo, Tokyo 113-0033, Japan}

\author{T. Yoshida}
\affiliation{Department of Physics and Department of Complexity 
Science and Engineering, University of Tokyo, 
Hongo, Tokyo 113-0033, Japan}
\affiliation{JST, TRIP, Chiyoda, Tokyo 102-0075, Japan}

\author{A. Fujimori}
\affiliation{Department of Physics and Department of Complexity 
Science and Engineering, University of Tokyo, 
Hongo, Tokyo 113-0033, Japan}
\affiliation{JST, TRIP, Chiyoda, Tokyo 102-0075, Japan}

\author{Y. Kotani}
\affiliation{Photon Factory, Institute of Materials Structure Science,
High Energy Accelerator Research Organization (KEK), Tsukuba, Ibaraki 305-0801, Japan}

\author{M. Kubota}
\affiliation{Photon Factory, Institute of Materials Structure Science,
High Energy Accelerator Research Organization (KEK), Tsukuba, Ibaraki 305-0801, Japan}

\author{K. Ono}
\affiliation{Photon Factory, Institute of Materials Structure Science,
High Energy Accelerator Research Organization (KEK), Tsukuba, Ibaraki 305-0801, Japan}

\author{M. Yi}
\affiliation{Department of Physics, Applied Physics, and Stanford Synchrotron Radiation Laboratory,
Stanford University, Stanford, California 94305, U.S.A.}

\author{D. H. Lu}
\affiliation{Department of Physics, Applied Physics, and Stanford Synchrotron Radiation Laboratory,
Stanford University, Stanford, California 94305, U.S.A.}

\author{R. Moore}
\affiliation{Department of Physics, Applied Physics, and Stanford Synchrotron Radiation Laboratory,
Stanford University, Stanford, California 94305, U.S.A.}

\author{Z.-X. Shen}
\affiliation{Department of Physics, Applied Physics, and Stanford Synchrotron Radiation Laboratory,
Stanford University, Stanford, California 94305, U.S.A.}

\author{A. Iyo}
\affiliation{Nanoelectronic Research Institute, National Institute of Advanced Industrial Science and Technology (AIST), Tsukuba, Ibaraki 305-8568, Japan}
\affiliation{JST, TRIP, Chiyoda, Tokyo 102-0075, Japan}

\author{K. Kihou}
\affiliation{Nanoelectronic Research Institute, National Institute of Advanced Industrial Science and Technology (AIST), Tsukuba, Ibaraki 305-8568, Japan}
\affiliation{JST, TRIP, Chiyoda, Tokyo 102-0075, Japan}

\author{H. Kito}
\affiliation{Nanoelectronic Research Institute, National Institute of Advanced Industrial Science and Technology (AIST), Tsukuba, Ibaraki 305-8568, Japan}
\affiliation{JST, TRIP, Chiyoda, Tokyo 102-0075, Japan}

\author{H. Eisaki}
\affiliation{Nanoelectronic Research Institute, National Institute of Advanced Industrial Science and Technology (AIST), Tsukuba, Ibaraki 305-8568, Japan}
\affiliation{JST, TRIP, Chiyoda, Tokyo 102-0075, Japan}

\author{S. Shamoto}
\affiliation{Japan Atomic Energy Agency, Tokai, Ibaraki 319-1195, Japan}
\affiliation{JST, TRIP, Chiyoda, Tokyo 102-0075, Japan}

\author{R. Arita}
\affiliation{Department of Applied Physics, University of Tokyo, Hongo, Tokyo 113-8656, Japan}
\affiliation{JST, TRIP, Chiyoda, Tokyo 102-0075, Japan}
\affiliation{JST, CREST, Hongo, Tokyo 113-8656, Japan}

\date{\today}

\begin{abstract}
We have performed an angle-resolved photoemission spectroscopy (ARPES)
study of the iron-based superconductor PrFeAsO$_{0.7}$ and examined the
Fermi surfaces and band dispersions near the Fermi level.  Heavily
hole-doped electronic states have been observed due to the polar nature
of the cleaved surfaces.  Nevertheless, we have found that the ARPES
spectra basically agree with band dispersions calculated in the local
density approximation (LDA) if the bandwidth is reduced by a factor of
$\sim$2.5 and then the chemical potential is lowered by $\sim$70 meV.
Comparison with previous ARPES results on LaFePO reveals that the energy
positions of the $d_{3z^2-r^2}$- and $d_{yz,zx}$-derived bands are
considerably different between the two materials, which we attribute to
the different pnictogen height as predicted by the LDA calculation.
\end{abstract}

\maketitle

The recent discovery of superconductivity in iron pnictides
\cite{ref:kamihara_As} has attracted keen attention in the materials
science community from both experimental and theoretical points of view
because they are the only class of superconductors which show critical
temperatures ($T_c$) reaching $\sim56$ K \cite{ref:wang} other than the
cuprates.  This new class of iron-based systems share some common
properties with the cuprates such as layered crystal structures
\cite{ref:kamihara_As} and antiferromagnetic ordering in the parent
compounds \cite{ref:cruz,ref:huang}.  However, many differences exist
between the two families especially in their electronic structures.
These differences started to appear from the early stage when
local-density-approximation (LDA) band-structure calculations predicted
that many Fe 3$d$-derived bands cross the Fermi level ($E_F$), resulting
in complicated hole- and electron-like Fermi surfaces (FS's)
\cite{ref:kuroki,ref:mazin,ref:ishibashi}, whereas only a single band
with one FS exists in the cuprates. The predictions of the LDA
calculations were confirmed by photoemission experiments, which
demonstrated that Fe 3$d$ states are predominant near $E_F$
\cite{ref:sato,ref:malaeb,ref:koizsch} with moderate \textsl{p-d}
hybridization and electron correlations \cite{ref:malaeb}.  Moreover,
angle-resolved photoemission spectroscopy (ARPES) studies have revealed
(i) several disconnected hole- and electron-like FS sheets
\cite{ref:liu101}, (ii) moderately renormalized energy bands due to
electron correlations \cite{ref:yang,ref:dhlu}, (iii) kinks in the
dispersions, suggesting coupling of quasiparticles to boson excitations
\cite{ref:richard,ref:wray}, and (iv) FS-dependent nodeless,
nearly-isotropic superconducting gaps
\cite{ref:ding,ref:kondo,ref:terashima,ref:nakayama}.  Although the
mechanism of superconductivity has not been elucidated yet, there is a
remarkable correlation between the $T_c$ and the position of pnictogen
atoms relative to the Fe plane \cite{ref:chlee}.

The ARPES observations mentioned above were mainly obtained for the
so-called 122 system while only a few results have been reported for the
1111 system \cite{ref:dhlu,ref:kondo,ref:dhlu2,ref:hliu,ref:holder}
owing to the difficulty in obtaining high quality, sizable single
crystals.  Because the 1111 system has higher $T_c$'s than those of the
122 system, detailed knowledge of their electronic structure
and their differences from that of the 122 system may give a clue
for understanding the mechanism of superconductivity in the iron-based
superconductors.  In the previous ARPES studies on the 1111
system~\cite{ref:dhlu,ref:dhlu2}, heavily hole-doped electronic states
have been observed probably due to the polar nature of the cleaved
surfaces.

In this work, we have performed ARPES measurements on the 1111 system
PrFeAsO$_{0.7}$, which has a $T_c$ as high as $\sim42$ K, and compared
the results with a band-structure calculation.  We have found that the
ARPES spectra agree well with the LDA band dispersions and FS's if the
calculated bandwidth is reduced by a factor of 2.5 and then the chemical
potential is lowered by 70 meV, resulting in heavily hole-doped
correlated electronic states.  We have thus found remarkable differences
in the electronic structures of PrFeAsO$_{0.7}$ and LaFePO
\cite{ref:dhlu}, which can be attributed to the change of the pnictogen
height, the distance between the Fe plane and the pnictogen atoms, as
predicted by a band structure calculation \cite{ref:vildosola}.

High-quality single crystals of the electron-doped compound
PrFeAsO$_{0.7}$ ($T_c\sim42$ K) were synthesized by a high-pressure
method described in Ref. \cite{ref:ishikado}.  The ARPES measurements
were carried out at BL5-4 of Stanford Synchrotron Radiation Laboratory
(SSRL), at BL10.0.1 of Advanced Light Source (ALS), and at BL-28A of
Photon Factory (PF) using incident photons of $h\nu=25$ eV
linearly-polarized, $h\nu=42.5$ eV linearly-polarized, and
$h\nu=36\text{-}80$ eV circularly-polarized , respectively.  SCIENTA
R4000 analyzers were used at SSRL and ALS and a SCIENTA SES-2002
analyzer was used at PF, with a total energy resolution of $\sim$15
meV and a momentum resolution of $\sim0.02~\pi/a$, where $a=4.0$ \AA ~is
the in-plane lattice constant.  The crystals were cleaved \textit{in
situ} at $T=10$ K in an ultra-high vacuum better than $1\times10^{-10}$
Torr giving flat mirror-like surfaces which stayed clean all over our
measuring time ($\sim2$ days).  The calibration of $E_{F}$ of the
samples was achieved by referring to that of gold which was in
electrical connect with the samples.  In the measurements at ALS, the
electric field vector $\bm E$ of incident photons was in the Fe plane
and its direction along the Fe-Fe nearest-neighbor.  In the measurements
at SSRL, although the direction of the in-plane component of $\bm E$ was
the same as that at ALS, $\bm E$ had a component perpendicular to the
plane.  We have performed a density functional calculation within the
LDA by using a WIEN2k package \cite{ref:blaha}, where the experimental
tetragonal lattice parameters of PrFeAsO at room temperature were used.
As for Pr 4{\it f} bands, we have adopted the LSDA+U method in order to
remove the bands away from $E_F$.

\begin{figure}[]
\begin{center}
\includegraphics[clip,width=8.5cm,trim=0 120 0 0]{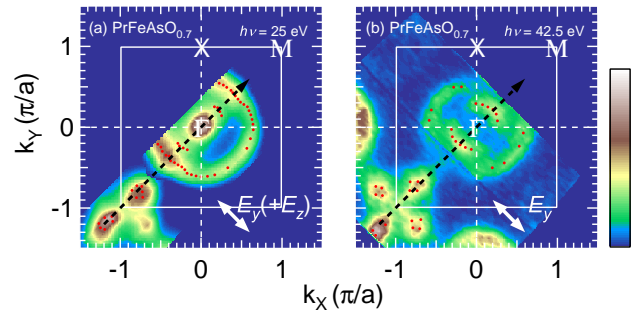}
\caption{(Color online) Fermi surface mapping of PrFeAsO$_{0.7}$
obtained by integrating the EDCs over an energy window of $E_F\pm5$ meV.
The white square highlights the boundary of the first Brillouin zone.
$a$ is the in-plane lattice constant.  The direction of polarization
vector is indicated in each panels.  Red dots indicate $k_F$ positions
determined by the peak positions of momentum distribution curves
(MDC's).  (a) Fermi surface mapping taken with $h\nu=25$ eV.  (b) Fermi
surface mapping taken with $h\nu=42.5$ eV.}\label{fig:FS}
\end{center}
\end{figure}

Figures \ref{fig:FS} (a) and \ref{fig:FS}(b) show the results of FS
mapping for the PrFeAsO$_{0.7}$ sample at low temperature ($\sim10$ K)
using photon energies $h\nu=25$ eV and $42.5$ eV, respectively.  Here,
we choose the local coordinate system around the Fe atom such that the
$x$ and $y$ axes point toward the nearest-neighbor Fe atoms.  The $x$
and $y$ direction are indicated by the electric field vector in Fig
\ref{fig:FS}.  In these plots, the photoemission intensity has been
integrated over $E_F\pm5$ meV.  In both plots one can clearly observe a
large nearly circular hole pocket with $k_F\sim0.6$($\pi/a$) centered at
the $\Gamma$ point of the the two-dimensional (2D) Brillouin zone (BZ).
A smaller nearly circular hole pocket with $k_F\sim0.3$($\pi/a$) is also
seen in Fig. \ref{fig:FS} (b) while the intensity is very weak in
Fig. \ref{fig:FS} (a).  In Fig. \ref{fig:FS} (b), the momentum regions
with strong intensities are opposite between the large and small FS
sheets around the $\Gamma$ point, implying that they have different
orbital characters.  The large size of the hole pocket has been reported
by the previous ARPES studies for the 1111 iron-based superconductors
\cite{ref:dhlu,ref:kondo,ref:dhlu2,ref:hliu} and reflects heavily
hole-doped electronic states.  One can also observe clover-shaped FS's
around the corner (the M point) of the 2D BZ.  This occurs because the
Fermi level is lowered below the four Dirac points around M caused by
excess hole doping.  The excess hole doping occurs because the cleaved
surface in the 1111 iron pnictides is electronically polar and
electronic charges must reconstruct after cleaving \cite{ref:tsukada}.
Note that the heavily hole-doped electronic states in the surface region
have been observed in spite of the fact the oxygen deficiency of the
bulk samples produces negative carrier.  The clover-shaped FS's have
been also observed in previous ARPES studies of KFe$_2$As$_2$
\cite{ref:K122_sato,ref:K122_yoshida}.

\begin{figure}[]
\begin{center}
\includegraphics[clip,width=7cm]{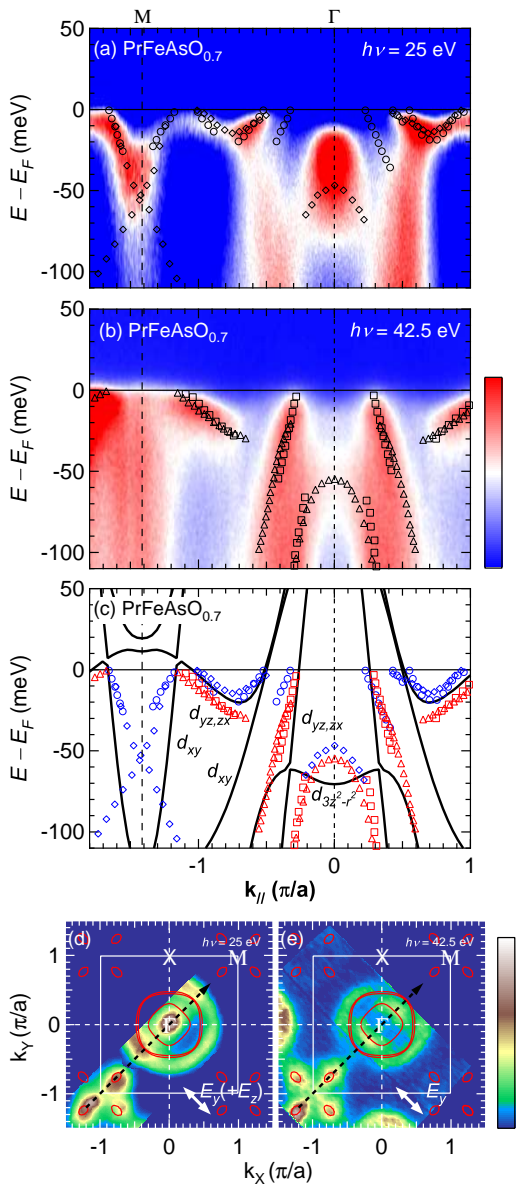} \caption{(Color online)
Comparison between the ARPES spectra of PrFeAsO$_{0.7}$ and the LDA band
structure along the $\Gamma$-M direction.  The directions of momentum
and polarization vector of (a) and (b) are indicated in
Fig. \ref{fig:FS}.  (a) The ARPES data taken with $h\nu=25$ eV.
Experimental band structure deduced from the second derivative plots of
EDCs and MDCs is also shown.  Diamond: EDC peak positions for $h\nu=25$
eV, circle: MDC peak positions for $h\nu=25$ eV.  (b) The same as (a)
with $h\nu=42.5$ eV.  Triangle: EDC peak positions for $h\nu=42.5$ eV,
square: MDC peak positions for $h\nu=42.5$ eV.  (c) Experimental band
structure of PrFeAsO$_{0.7}$ deduced from (a) and (b).  The calculated
band dispersions are plotted after reducing the bandwidth by a factor of
2.5 and then shifting down the chemical potential by 70 meV.  (d)
Calculated Fermi surface plotted on the Fermi surface mapping taken with
$h\nu=25$ eV.  (e) The same as (d) with $h\nu=42.5$ eV.}\label{fig:Ek}
\end{center}
\end{figure}

The ARPES intensity plots in energy-momentum ($E$-$k$) space along the
$\Gamma$-M direction taken at $h\nu=25$ eV and $42.5$ eV are shown in panels (a)
and (b) of Fig. \ref{fig:Ek}, respectively.  The direction of the
electrical polarization vector of incident light is indicated in
Fig. \ref{fig:FS}.  In Fig. \ref{fig:Ek} (a) and (b), the band
dispersions of PrFeAsO$_{0.7}$ deduced from the second derivative plots
of EDC's and those of MDC's are also shown (see caption).  In order to
understand the multiband electronic structure of this material, we plot
the experimentally deduced band structure and compare with the LDA
dispersions in Fig. \ref{fig:Ek} (c).  We have found that the band
structure basically agree with the calculated band dispersions if the
bandwidth is reduced by a factor of 2.5 and then the chemical potential
is lowered by 70 meV.  The band narrowing is due to electron correlations
which are not taken into account in the LDA calculation and the chemical
potential shift is due to the electronic reconstruction of the surface
layers to prevent the ``polar catastrophe'' \cite{ref:tsukada}.  One can
also reproduce the observed FS's using the same shift as shown in
Fig. \ref{fig:Ek} (d) and \ref{fig:Ek} (e).

\begin{figure*}[]
\begin{center}
\includegraphics[clip,width=17cm]{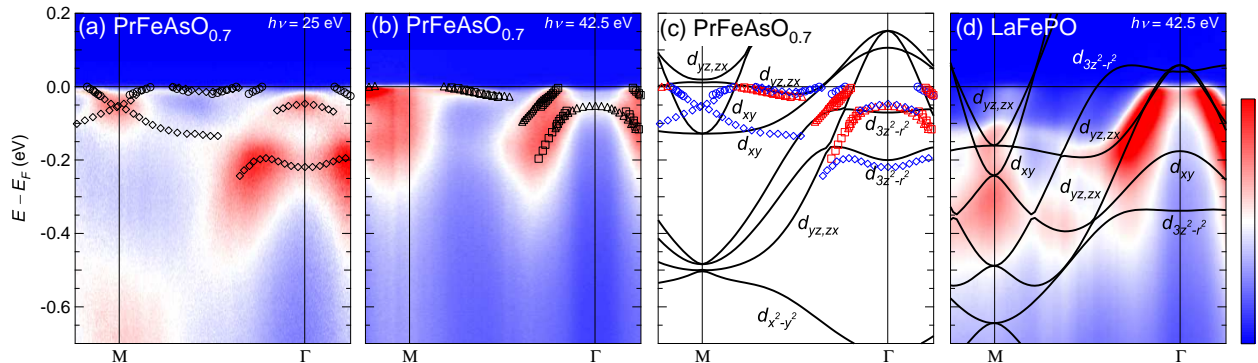} \caption{(Color online)
Comparison between ARPES spectra and LDA band structures along the
$\Gamma$-M direction.  The directions of momentum and polarization
vector of (a) and (b) are indicated in Fig. \ref{fig:FS}.  (a) The same
as Fig. \ref{fig:Ek} (a).  (b) The same as Fig. \ref{fig:Ek} (b).  (c)
The same as Fig. \ref{fig:Ek} (c). (d) The ARPES data of LaFePO taken
from Ref. \cite{ref:dhlu}.  }\label{fig:wide}
\end{center}
\end{figure*}

According to the LDA calculation [Fig. \ref{fig:Ek} (c)], the outer two
FS's around the $\Gamma$ point are nearly degenerate and consist of
$d_{yz,zx}$ and $d_{xy}$ bands while the inner one consists of only a
$d_{yz,zx}$ band.  Since the spectral intensity of the $d_{xy}$ band
would be weak and may not be seen near the $\Gamma$ point due to matrix
element effects \cite{ref:yzhang}, one can conclude that the intensities
of both the outer and inner FS's in Figs. \ref{fig:FS} (b) and
\ref{fig:Ek} (b) mainly come from the $d_{yz,zx}$ bands.  Furthermore,
if we take into account matrix element effects for the electric vector
$\bm{E}$ in Fig. \ref{fig:FS} (b), the outer and inner FS's have
$d_{zx}$ ($d_{yz}$) and $d_{yz}$ ($d_{zx}$) orbital character in the
$k_x$ ($k_y$) direction, respectively \footnote{The data of
Figs. \ref{fig:FS} (b), \ref{fig:Ek} (b), and \ref{fig:wide} (b) have
been taken with $\sigma$ geometry \cite{ref:yzhang}.  Therefore, the
spectral intensity of $d_{zx}$ band cannot be observed.}.  However, the
band-structure calculation predicts opposite orbital characters between
them, namely, $d_{yz}$ ($d_{zx}$) character for the outer FS and
$d_{zx}$ ($d_{yz}$) character for the inner one along the $k_x$ ($k_y$)
direction \cite{ref:gracer}.  Although the origin of the discrepancy
between experiment and calculation is not clear at present, a similar
discrepancy has been reported in a previous ARPES results on a 122
iron-based superconductor~\cite{ref:yzhang}.

Here, we shall discuss the orbital character of the other bands seen in
Figs. \ref{fig:Ek} (a) and (b).  As for the band observed $\sim$50 meV
below $E_F$ around the $\Gamma$ point, one can notice that the spectral
intensity in Fig. \ref{fig:Ek} (a), where the $E_z$ component is finite,
is strong compared to that in Fig. \ref{fig:Ek} (b).  Therefore, this
band is considered to have $d_{3z^2-r^2}$ character as predicted by the
band-structure calculation.  From Fig. \ref{fig:Ek} (c), one can also
see that the FS around the M point has a hole-like feature arising from
the intersection of the $d_{yz,zx}$ band and the $d_{xy}$ band near
$E_F$.

Calculation of the volume enclosed by the hole FS's yields hole counts
of 0.07, 0.28 and 0.05 per Fe atom for the inner FS around the
$\Gamma$ point, the outer FS around the $\Gamma$ point, and the
FS's around the M point, respectively.
As mentioned above, the spectral intensity
of the $d_{xy}$ band should be weak and cannot be observed near the
$\Gamma$ point and, therefore, we cannot evaluate the size of the $d_{xy}$
band FS.  For three possible cases, (1) the $d_{xy}$ FS has the
same size as the outer FS, (2) it has the same size as the inner FS, and
(3) it does not exist, the total hole concentration
becomes (1) 0.68, (2) 0.47, and (3) 0.40 holes per Fe atom, respectively.  These
values are comparable to the predicted value of 0.5 which is necessary
to avoid the polar catastrophe \cite{ref:tsukada}.

Now, let us compare the present results with the previous ARPES results
on LaFePO with $T_c\sim$6 K \cite{ref:kamihara_P}, which also has the
same structure as PrFeAsO with lower pnictogen height than that of
PrFeAsO$_{0.7}$. A band structure calculation \cite{ref:vildosola}
predicts that if the pnictogen height is lowered, the $d_{3z^2-r^2}$
band is raised and may cross $E_F$ around the $\Gamma$ point.  In order
to see differences in the electronic structure between PrFeAsO$_{0.7}$
and LaFePO, in Fig. \ref{fig:wide} we compare the present ARPES
intensity $E$-$k$ plot with that of LaFePO \cite{ref:dhlu} along the
$\Gamma$-M line.  For the same reason as mentioned above,
PrFeAsO$_{0.7}$ has two bands at the $\Gamma$ point $\sim$50 meV and
$\sim$0.2 eV below $E_F$ with $d_{3z^2-r^2}$ character as predicted by
the band-structure calculation [Fig. \ref{fig:wide} (c)].  In LaFePO, in
contrast to PrFeAsO$_{0.7}$, one of the $d_{3z^2-r^2}$ bands crosses
$E_F$ and forms a quite large hole FS as shown in Fig. \ref{fig:wide}
(d).  In addition, in LaFePO, the $d_{yz,zx}$ bands around the M point
are lowered compared to those in PrFeAsO$_{0.7}$ and recover the
electron FS's.

Although we have not been able to observe the spectral intensity of the
$d_{xy}$ band near the $\Gamma$ point as mentioned above, it seems from
comparison between the data and band-structure calculation
[Figs. \ref{fig:wide} (c) and (d)] that PrFeAsO$_{0.7}$ has a $d_{xy}$
FS around the $\Gamma$ point while LaFePO does not.  According to the
theory of spin-flucutuation-mediated superconductibity \cite{ref:KKPRB},
in which the $d_{xy}$ FS plays an important role to induce high $T_c$
superconductivity, this may be the main reason why the $T_c$ of
PrFeAsO$_{0.7}$ is higher than that of LaFePO.

In a previous ARPES study of another 1111 superconductor LaFeAsO
\cite{ref:LXY}, the Dirac points around the M point are below $E_F$ like
in LaFePO [Fig. \ref{fig:wide} (d)], while they are slightly above $E_F$ in
PrFeAsO$_{0.7}$ [Figs. \ref{fig:Ek} (c) and \ref{fig:wide} (c)].  This
difference can also be explained by the change of the different
pnictogen heights based on band-structure calculation.

In summary, we have performed an ARPES study of the iron-based
superconductor PrFeAsO$_{0.7}$ and revealed the FS's and band
dispersions near $E_F$.  Although heavily hole-doped electronic states
have been observed, we have found that the ARPES spectra basically agree
with the calculated band dispersions if the bandwidth is reduced by a
factor of 2.5 and then the chemical potential is lowered by 70 meV.
This observation confirms that the LDA calculations for 1111 iron
pnictides capture the electronic structure in those compounds.  From the
comparison of the electronic structures between PrFeAsO$_{0.7}$ and
LaFePO, we have demonstrated the pnictogen height dependence of the
electronic structure in the 1111 pnictide series as predicted by the
band-structure calculations.

The authors acknowledge T. Hanaguri for informative discussions.  ALS is
operated by the Department of Energy (DOE) Office of Basic Energy
Science, Division of Materials Science, under Contract
No. DE-AC02-05CH11231.  SSRL is operated by the DOE Office of Basic
Energy Science Divisions of Chemical Sciences and Material Sciences.
Experiment at Photon Factory was approved by the Photon Factory Program
Advisory Committee (Proposal No. 2009S2-005).  Illuminating discussions
at A3 Foresight Program are gratefully acknowledzed.

\end{document}